\documentclass[usenatbib,usedcolumn,usegraphicx]{mn2e}
\begin{document}

\newcommand{\kms}{km\ s$^{-1}$}
\newcommand{\kmsmpc}{km\ s$^{-1}$\ Mpc$^{-1}$}
\newcommand{\ergscm}{erg\ s$^{-1}$ cm$^{-2}$} 
\newcommand{\ergcms}{erg\ s$^{-1}$ cm$^{-2}$} 
\newcommand{\OII}{\left[O_{\ II}\right]}
\newcommand{\Deg}{^{\circ}}
\newcommand{\phm}{\phantom}

\newcommand\aj{AJ} 
\newcommand\araa{ARA\&A} 
\newcommand\apj{ApJ} 
\newcommand\apjl{ApJ} 
\newcommand\apjs{ApJS} 
\newcommand\ao{Appl.~Opt.} 
\newcommand\apss{Ap\&SS} 
\newcommand\aap{A\&A} 
\newcommand\aapr{A\&A~Rev.} 
\newcommand\aaps{A\&AS} 
\newcommand\azh{AZh} 
\newcommand\baas{BAAS} 
\newcommand\jrasc{JRASC} 
\newcommand\memras{MmRAS} 
\newcommand\mnras{MNRAS} 
\newcommand\pra{Phys.~Rev.~A} 
\newcommand\prb{Phys.~Rev.~B} 
\newcommand\prc{Phys.~Rev.~C} 
\newcommand\prd{Phys.~Rev.~D} 
\newcommand\pre{Phys.~Rev.~E} 
\newcommand\prl{Phys.~Rev.~Lett.} 
\newcommand\pasp{PASP} 
\newcommand\pasj{PASJ} 
\newcommand\qjras{QJRAS} 
\newcommand\skytel{S\&T} 
\newcommand\solphys{Sol.~Phys.} 
\newcommand\sovast{Soviet~Ast.} 
\newcommand\ssr{Space~Sci.~Rev.} 
\newcommand\zap{ZAp} 
\newcommand\nat{Nature} 
\newcommand\iaucirc{IAU~Circ.} 
\newcommand\aplett{Astrophys.~Lett.} 
\newcommand\apspr{Astrophys.~Space~Phys.~Res.} 
\newcommand\bain{Bull.~Astron.~Inst.~Netherlands} 
\newcommand\fcp{Fund.~Cosmic~Phys.} 
\newcommand\gca{Geochim.~Cosmochim.~Acta} 
\newcommand\grl{Geophys.~Res.~Lett.} 
\newcommand\jcp{J.~Chem.~Phys.} 
\newcommand\jcap{J. Cosmology Astropart. Phys.}
\newcommand\jgr{J.~Geophys.~Res.} 
\newcommand\jqsrt{J.~Quant.~Spec.~Radiat.~Transf.} 
\newcommand\memsai{Mem.~Soc.~Astron.~Italiana} 
\newcommand\na{New Astronomy} 
\newcommand\nphysa{Nucl.~Phys.~A} 
\newcommand\physrep{Phys.~Rep.} 
\newcommand\physscr{Phys.~Scr} 
\newcommand\planss{Planet.~Space~Sci.} 
\newcommand\procspie{Proc.~SPIE} 

\title{Radio Monitoring Campaigns of Six Strongly Lensed Quasars}


\author[N. Rumbaugh et al.]{
N. Rumbaugh$^1$, 
C. D. Fassnacht$^1$, 
J. P. McKean$^{2,3}$, 
L. V. E. Koopmans$^3$,\newauthor
M. W. Auger$^4$, 
S. H. Suyu$^5$\\
$^1$Department of Physics, University of California, Davis, 1 Shields Avenue, Davis CA 95616, USA\\
$^2$ASTRON, The Netherlands Institute for Radio Astronomy, Postbus 2, NL-7990 AA Dwingeloo, The Netherlands \\
$^3$Kapteyn Astronomical Institute, University of Groningen, P.O. Box 800, 9700-AV Groningen, The Netherlands \\
$^4$Institute of Astronomy, Madingley Road, Cambridge CB3 0HA, UK \\
$^5$Institute of Astronomy and Astrophysics, Academia Sinica (ASIAA), P.O. Box 23-141, Taipei 10617, Taiwan 
}




\maketitle

\begin{abstract}

We observed six strongly lensed, radio-loud quasars (MG 0414+0534, CLASS B0712+472, JVAS B1030+074, CLASS B1127+385, CLASS B1152+199, and JVAS B1938+666) in order to identify systems suitable for measuring cosmological parameters using time delays between their multiple images. These systems are in standard two and four image configurations, with B1938 having a faint secondary pair of images. Two separate monitoring campaigns were carried out using the VLA and upgraded JVLA.  Lightcurves were extracted for each individual lensed image and analyzed for signs of intrinsic variability. While it was not possible to measure time delays from these data, $\chi^2$-based and structure function tests found evidence for variability in a majority of the lightcurves. B0712 and B1030 had particularly strong variations, exhibiting linear flux trends. These results show that most of these systems should be targeted with followup monitoring campaigns, especially B0712 and B1030.


\end{abstract}


\section{Introduction}

With cosmology entering an era of precision, the bulk of observational evidence has come to support the standard $\Lambda$CDM model with negligible curvature dominated by dark energy and dark matter \citep[see e.g.,][]{hinshaw13,planck13}. However, recent studies have yielded potentially discrepant values of $H_0$ \citep[see e.g.,][]{reid10,riess11,planck13}. For example, \citet{riess11} find $H_0 = 73.8\pm2.4$\ km s$^{-1}$ Mpc$^{-1}$, while the results from the Planck satellite, in conjunction with the Wilkinson Microwave Anisotropy Probe polarization data, yield $H_0 = 67.3\pm1.2$\ km s$^{-1}$ Mpc$^{-1}$ \citep{planck13}, which are in tension at the 2.4$\sigma$ level\footnote{See \citet{ef14} for a possible revision of the cosmological parameters measured by \citet{riess11}.}. While the tension could be due to systematics, the conflicts in cosmological parameters are not currently at a level necessary to conclusively rule out the $\Lambda$CDM model\footnote{Some extensions of the standard model have been proposed to reconcile the cosmological parameter tensions, such as the addition of an additional, sterile neutrino \citep[See e.g.,][]{HH13,BM14,wyman14}.}; more precise measurements are still required. 

While Planck data have provided a wealth of cosmological information, observations of the Cosmic Microwave Background provide mostly indirect information on the present era. Thus, lower redshift measurements can alleviate parameter degeneracies and accurately measure $H_0$. While distance ladder techniques have been a prominent tool since the establishment of cosmology \citep[e.g.,][]{riess11,free12}, additional independent methods will only help the community. 


Gravitational lens systems provide one such a method. \citet{ref64} first proposed using time delays between the lightcurves of variable strongly lensed images to measure angular diameter distances, and, thus, $H_0$ and other parameters. Early attempts using strongly lensed quasars suffered from a number of problems, including inadequate lightcurves, poor lens reconstruction, and insufficient attempts to overcome the mass sheet and density profile degeneracies \citep[See e.g.,][]{lehar92,fass98,courbin03,KS04}. Significant advances in lens modeling techniques and time delay measurement have brought strong lensing to a level comparable with the more prominent techniques of cosmological parameter inference. Recently \citet{suyu13} made a robust inference of $H_0$ using this technique through a detailed analysis of the CLASS B1608+656 and RX J1131-1231 systems (hereafter B1608 and J1131). The study used comprehensive observations of both lensed quasar systems and their environments. One of the benefits of this method is that every lens system provides an independent and relatively high precision measurement of the cosmological parameters, so increasing the sample size by a factor of $N$ reduces random uncertainties on the cosmological parameter measurements by a factor of roughly $\sqrt{N}$. 


Here, we have sought to expand the sample of strongly lensed quasars with which useful measurements of cosmological parameters can be made. In order to be used for time delay measurements, the flux of the lensed quasar must exhibit measurable intrinsic variability that is distinguishable from sources of extrinsic variability such as galactic scintillation \citep[See e.g.,][]{koop03}. In this paper we present the results of a radio monitoring campaign of six lenses, observed with the NRAO\footnote{The National Radio Astronomy Observatory is a facility of the National Science Foundation operated under cooperative agreement by Associated Universities, Inc.} Very Large Array (VLA) and the Janksy Very Large Array (JVLA). Our choice of radio, as opposed to optical, monitoring campaigns has several motivations. First, while optical observations probe a smaller physical region of the lensed quasar that is more likely to be variable, this also makes them more vulnerable to false signals from microlensing, as in the first several seasons of observations of J1131 by \citet{tewes13}. The nature of radio observations also allow for 24-hour and year-round observation, with minimal impacts by weather, which aids monitoring. Finally, radio observations allow monitoring of systems where optical monitoring would be difficult to impossible because emission from the lensing galaxy overpowers the optical output of the lensed active galactic nucleus.

In this paper, we will first discuss our sample of lenses in Section \ref{sec:lens}. In Section \ref{sec:obs}, we review the details of the observations. In Section \ref{sec:res}, we present and analyze our results, including our variability tests and, where possible, measurements of time delays. In Section \ref{sec:dis}, we discuss the results and their implications for our lenses and the feasibility of using them to measure time delays. 

\begin{table}
\caption{Lensed System Details}
\label{LensTab}
\begin{tabular}{lcccc}
\hline
\footnotesize{Lens}
 & \footnotesize{Num. of}
 & \footnotesize{Lens}
 & \footnotesize{Lens}
 & \footnotesize{Source}\\
\footnotesize{}
 & \footnotesize{Images}
 & \footnotesize{Config.}
 & \footnotesize{Redshift}
 & \footnotesize{Redshift}\\
\hline
MG0414 & 4\phm{+2} & fold\phm{db} & 0.96\phm{0} & 2.64\phm{0} \\
B0712  & 4\phm{+2} & fold\phm{db} & 0.40\phm{0} & 1.33\phm{0} \\
B1030  & 2\phm{+2} & double & 0.599 & 1.535 \\
B1127  & 2\phm{+2} & double & $^{b}$ & $^{b}$ \\
B1152  & 2\phm{+2} & fold\phm{db} & 0.439 & 1.019 \\
B1938  & 4+2       & fold$^{a}$\phm{d} & 0.88\phm{0} & 2.059 \\
\hline
\multicolumn{5}{p{7cm}}{$^{\rm a}${\footnotesize{B1938 contains four images in a fold geometry plus a fainter, separate doubly-lensed source.}}}\\
\multicolumn{5}{l}{$^{\rm b}${\footnotesize{No definitive redshift measured.}}}
\end{tabular}
\end{table}

\begin{figure*}
\begin{center}
\includegraphics[width=0.8\textwidth]{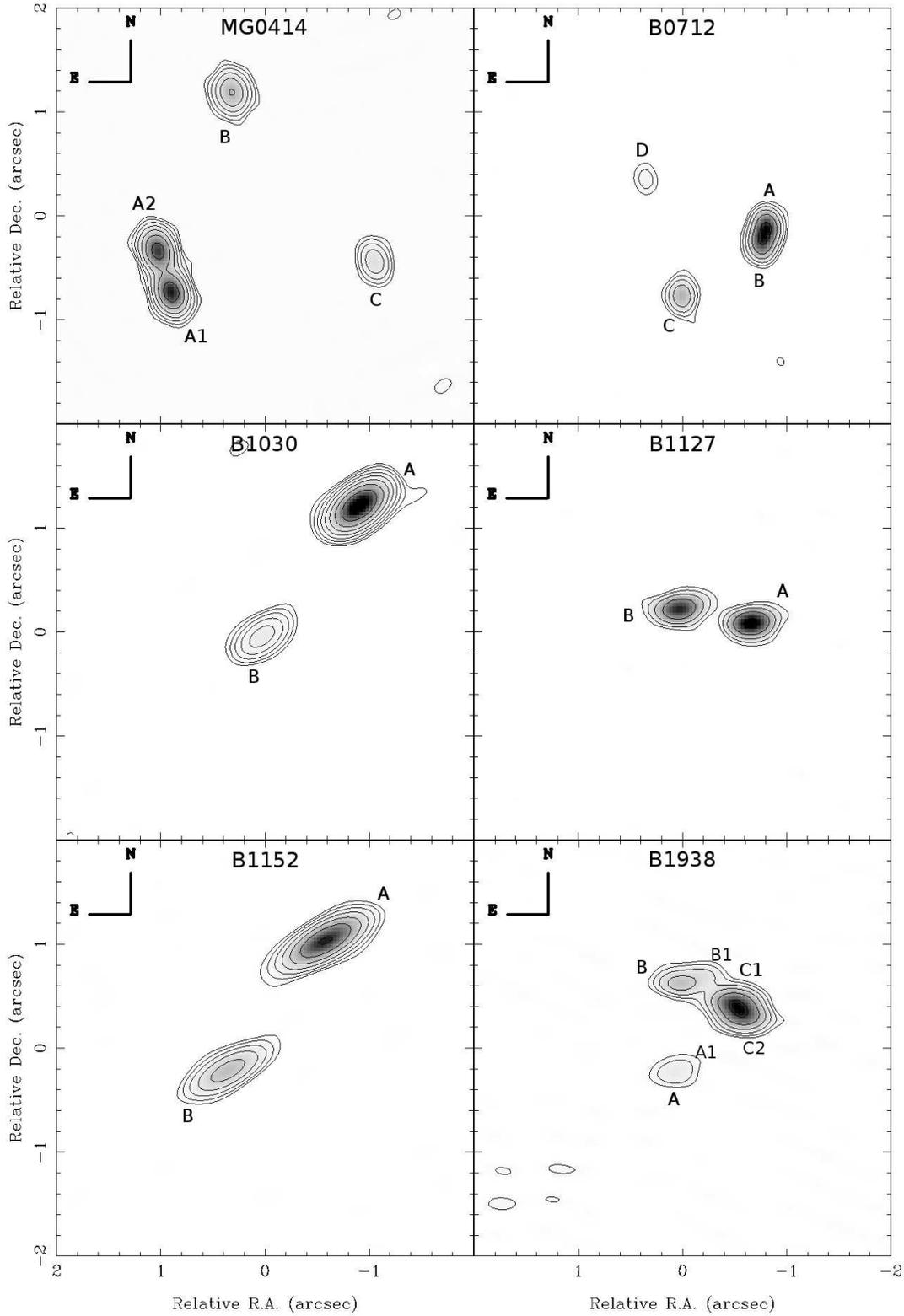}
\end{center}
\caption{8.46 GHz radio contours are shown for all lensed systems, with each image labeled. The maps are centered at the following coordinates: MG0414 - 04:14:37.7,+05:34:43.0; B0712 - 07:16:03.7,+47:08:50.3; B1030 - 10:33:34.1,+07:11:25.0; B1127 - 11:30:00.2,+38:12:03.1; B1152 - 11:55:18.3,+19:39:41.3; B1938 - 19:38:25.4,+66:48:52.6. All maps, except for B1938, were created from observations taken on October 25, 2000, while the VLA was in the A configuration. The B1938 map was created from observations taken on July 21, 2011, while the JVLA was in the A configuration. Contour levels are $3\times\left(2^i\right)$ times the RMS level, with the lowest contour at $i=1$. Note that B1938 has a 4+2 image configuration, with a pair of fainter images, A1 and B1. Slight extensions of the contours can be observed at their positions.}
\label{fig:radmaps}
\end{figure*}

\section{The Sample}
\label{sec:lens}

Six lenses are studied in this paper. These are MG 0414+0534, CLASS B0712+472, JVAS B1030+074, CLASS B1127+385, CLASS B1152+199, and B1938+666 (hereafter MG0414, B0712, B1030, B1127, B1152, and B1938, respectively). The radio images of each are shown in Figure \ref{fig:radmaps} and some of their properties are summarized in Table \ref{LensTab}. In this section, we will briefly discuss each system. 

\subsection{MG 0414+0534}

MG0414 was discovered as part of the MIT-Green Bank 5 GHz survey using the VLA \citep{ben86,hewitt92}. It consists of a lensing galaxy and four images of a background quasar in a fold lens geometry \citep{law95}. At radio wavelengths, the images have a maximum separation of $2\farcs1$, while the two brightest images are separated by only $\sim 0.5 ''$, and are thus blended to varying degrees in radio images \citep{hewitt92,katz93}. The lensing galaxy has a redshift of $z = 0.96$ and an extremely red color \citep{TK99}. Similarly, the quasar has a redshift of $z = 2.64$ and a reddened color \citep{law95,bar98}. The source images have different colors, suggesting some reddening occurs because of the dusty lens galaxy \citep{law95}. \citet{MH97} found the images to have root mean square variations (RMS) of $\sim 3.5$\% through VLA monitoring. Single dish monitoring found the total flux of the system to increase by 10-15\% over the course of 18 months \citep{cast11}. Modeling has predicted time delays of 12.5, 12.3, and 70.8 days for the time delays between image B and images A1, A2, and C, respectively (Moustakas et al., in prep.).

\subsection{CLASS B0712+472}

B0712 was discovered as part of the Cosmic Lens All-Sky Survey \citep[CLASS;][]{browne03,myers03} and consists of four images of a quasar in a fold lens geometry, with a maximum image separation of $1\farcs3$ \citep{jack98}. The two brightest images are separated by $\sim 0.2''$ and are thus highly blended in VLA imaging. Optical spectroscopy yielded a source redshift of $z = 1.33$ and a lens redshift of $z = 0.40$ \citep{FC98}. The total source magnitudes are $V \sim 23$ and $I \sim 22.5$, assuming a point source \citep{FC98}. These optical source flux densities significantly differed for observations separated by periods on the order of months, which could indicate the system is variable due to intrinsic variations of the lensed sources, or because of microlensing by stars within the lens \citep{FC98}. Modeling has predicted time delays of 9 days for the time delay between image C and the close images A and B and a time delay of 20.4 days between image C and image D (Moustakas et al., in prep.).

\subsection{JVAS B1030+074}

B1030 was discovered as part of the Jodrell Bank-VLA Astrometric Survey \citep[JVAS;][]{pat92}; \citep{xan98}. The system is composed of two flat-spectrum radio images separated by $1\farcs6$ and with a flux ratio of 15:1 \citep{xan98}. The brighter image A has measured magnitudes of $V \sim 20$ and $I \sim 19$, while image B is approximately 2.5 to 3 magnitudes fainter in both bands. \citet{FC98} measure the redshifts of the lensing galaxy and source to be $z = 0.599$ and $z = 1.535$, respectively. The lens was determined to consist of a main and a satellite galaxy \citep{xan98,lehar00}. \citet{xan98} predict a time delay of 156/$h_{50}$ days (111 days with $H_0 = 70$\kmsmpc), while \citet{saha06}, using $H_0 = 70$\kmsmpc, predict 153$^{+29}_{-57}$ days.

\subsection{CLASS B1127+385}

B1127 was discovered as part of CLASS. The system is composed of two images separated by $\sim 0.7$ arcseconds, with flat radio spectra \citep{koop99}. Hubble Space Telescope (HST) imaging shows two lens galaxies, with likely redshifts of $z \ge 0.5$, and mass modeling supports a double lens \citep{koop99}. No definitive redshift has been measured for either the source or lensing galaxy. While this would hinder the extraction of cosmological information from this system, observations of variability in the system would justify further attempts at a redshift measurement.

\subsection{CLASS B1152+199}

B1152 was also discovered as part of CLASS. The system consists of two flat-spectrum images separated by $1\farcs6$ with a flux ratio of $\sim$ 3:1 \citep{myers99}. The lens and source have measured redshifts of $z = 0.439$ and $z = 1.019$, respectively \citep{myers99}. A third and fourth radio source were detected $\sim$ 20-40$''$ from the others on either side, and could be radio lobes from either the lensing or background galaxies \citep{myers99}. \citet{rusin02} predict a time delay of $35.9\pm 2$ days when using an isothermal sphere mass model with $H_0 = 100$\kmsmpc.

\subsection{JVAS B1938+666}

B1938 was discovered as part of JVAS. Initial radio observations showed four flat-spectrum components with a maximum separation of $0\farcs95$ in a fold geometry, as well as a second, much fainter doubly lensed source. \citep{pat92,rhoads96,king97}. Later observations revealed a very red source galaxy along with a nearly complete infrared Einstein ring \citep{rhoads96,king98,lag12}. The lensing galaxy and primary source have measured redshifts of $z = 0.88$ and $z = 2.059$, respectively \citep{TK00,rie11}. 

\begin{table*}
\caption{Observation Details}
\label{DatesTab}
\begin{tabular}{lcccccc}
\hline
\footnotesize{Lens}
 & \footnotesize{Campaign}
 & \footnotesize{Campaign}
 & \footnotesize{$N_{obs}$$^{\rm a}$}
 & \footnotesize{$N_{good}$$^{\rm b}$}
 & \footnotesize{Average}
 & \footnotesize{Median Obs.}\\
\footnotesize{}
 & \footnotesize{Start}
 & \footnotesize{End}
 & \footnotesize{}
 & \footnotesize{}
 & \footnotesize{Spacing (Days)$^{\rm c}$}
 & \footnotesize{Length (sec)}\\
\hline
\multicolumn{7}{l}{VLA Campaign} \\
\hline
MG0414 & 10/2000 & 5/2001 & 63 & 45 & 3.5(4.9) & 160 \\
B0712 & 10/2000 & 5/2001 & 63 & 50 & 3.5(4.4) & 420 \\
B1030 & 10/2000 & 5/2001 & 63 & 52 & 3.5(4.3) & 190 \\
B1127 & 10/2000 & 5/2001 & 63 & 53 & 3.5(4.2) & 300 \\
B1152 & 10/2000 & 5/2001 & 63 & 50 & 3.5(4.4) & 310 \\
\hline
\multicolumn{7}{l}{JVLA Campaign} \\
\hline
MG0414 & 6/2011 & 9/2011 & 18 & 14 & 5.3(5.8) & 40 \\
B0712 & 6/2011 & 9/2011 & 18 & 14 & 5.3(5.8) & 215 \\
B1938 & 6/2011 & 9/2011 & 26 & 24 & 3.6(3.8) & 110 \\
\hline
\multicolumn{7}{l}{$^{\rm a}${\footnotesize{Total number of observations taken.}}}\\
\multicolumn{7}{l}{$^{\rm b}${\footnotesize{Total number of observations used in analysis.}}}\\
\multicolumn{7}{l}{$^{\rm c}${\footnotesize{Average spacing in days between subsequent observations, using all (only good) observations.}}}\\
\end{tabular}
\end{table*}

\begin{figure*}
\begin{center}
\includegraphics[width=\textwidth]{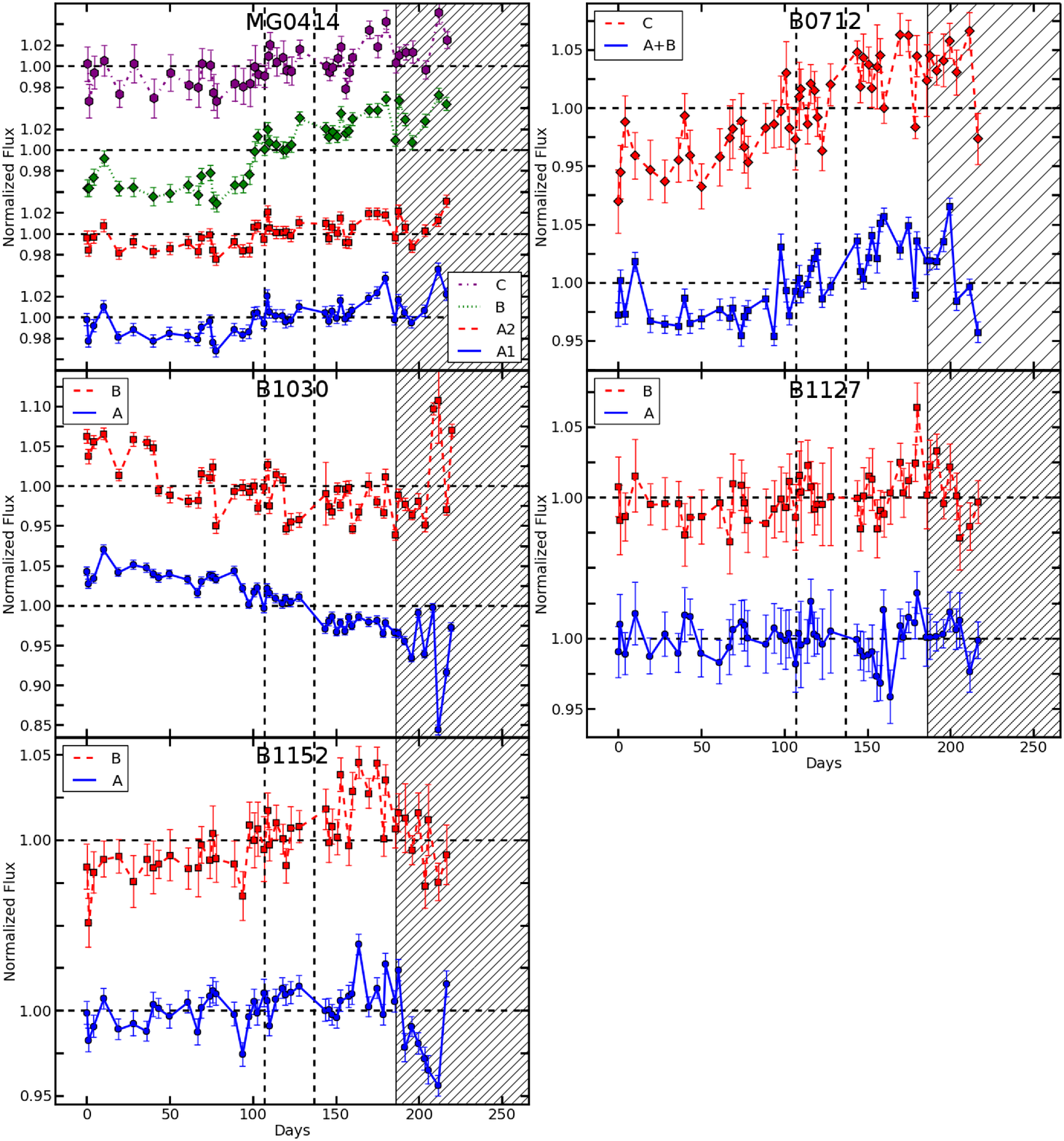}
\end{center}
\caption{
  Lightcurves for the five lensed systems observed using the VLA, with arbitrary vertical offsets between images in each panel. Each lightcurve was normalized by dividing by its mean. On the horizontal axis is shown days since October 21, 2000 and the vertical axis shows the normalized flux values around each curve. A horizontal dashed line indicates the mean of each image's lightcurve. In each case, problematic observations were removed. See Section \ref{sec:obs} for more details. The shaded region shows where all observations exhibited systematic variations that we were unable to remove. Enclosed data points were not used in analysis. See Section \ref{sec:var} for details. For B0712, the fluxes for images A and B were added together. This pair of images was highly blended, even in the A configuration. In addition, the low signal-to-noise lightcurve of the faint B0712 D image is omitted. Lightcurves for all other images are shown individually. The vertical lines show when the VLA switched from the A to the BnA configuration, and then from the BnA to the B configuration. Mean fluxes for the lightcurves are listed in Table \ref{VarTab}}
\label{fig:lightcurves_norm}
\end{figure*}

\section{Observations and Reduction}
\label{sec:obs}

We observed the radio sources with two different monitoring campaigns. Some information on the campaigns is summarized in Table \ref{DatesTab}. MG0414, B0712, B1030, B1127, and B1152 were observed at 8.46 GHz between November 2000 and May 2001 using the VLA in its A, BnA, and B configurations. These five lenses were observed approximately once every four days. Three compact symmetric objects (CSOs) were also observed as part of the program. One was used as the primary flux calibrator while the other two were secondary flux calibrators. CSOs are steep-spectrum radio sources with low variability and are shown to be stable flux calibrators at 8.5 GHz \citep{FT01}. 

MG0414, B0712, and B1938 were observed at 8.46 GHz between June and September 2011 on the JVLA using the A configuration. MG0414 and B0712 were observed together approximately once every 5.3 days, and B1938 was observed on average once every 4.1 days. Five to six CSOs were observed with each B1938 observation as secondary flux calibrators, while two were observed for the MG0414/B0712 block. The CSOs used were chosen according to the time of the observation. All observations used 3C48 as the primary flux calibrator, except for some cases for B1938, which used 3C286. Each observation of the lens(es) and calibrators was 30 minutes in duration.

The VLA observations were reduced, using the Astronomical Image Processing System\footnote{http://www.aoc.nrao.edu/aips/}, as in \citet{fass02}. The JVLA observations were reduced using the Common Astronomy Software Applications (CASA) package version 3.4. After flagging bad visibilities, the flux scales for the primary flux calibrators were set using the CASA task {\it setjy}. The tasks {\it bandpass} and {\it gaincal} were used to solve for the antenna-based delays, the complex bandpass, the amplitude gains, and the phase gains. The relative gains for all calibrators were then determined using {\it fluxscale}. 

To clean the data and calculate the fluxes of the different lens components, we used the Difmap software package \citep{shep94}. Point sources were fit simultaneously for the two or four brightest components of each system, depending on whether it was a double or quad configuration lens. The secondary pair of images in the B1938 system were included in our fitting, but their inclusion or exclusion in our models had no discernible effect on the apparent flux for the other images, and the faint pair of images had very low signal to noise ratios, and so the results for images A1 and B1 are not reported here.

Difmap was also used to extract fluxes for the CSOs. Since the CSOs are expected to have very low variability, any correlation in their fluxes can be interpreted as a variation in absolute flux calibration \citep{FT01}. The lightcurves of the lensed images were then divided by the average normalized lightcurves of the CSOs to remove non-intrinsic variation in the data. 

\begin{figure}
\begin{center}
\includegraphics[width=0.5\textwidth]{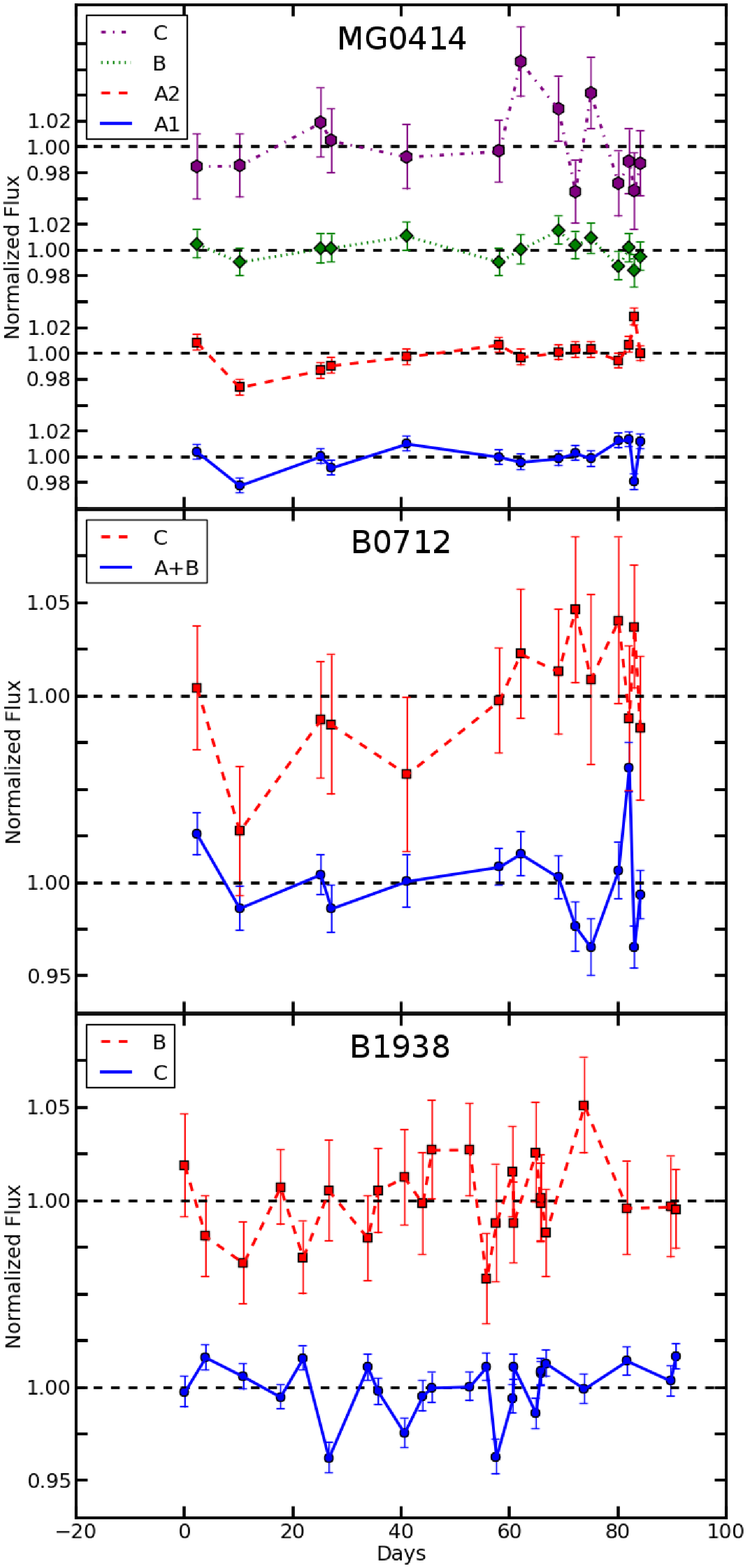}
\end{center}
\caption{
  Lightcurves for the three lensed systems observed using the JVLA, with arbitrary vertical offsets between images in each panel. Each lightcurve was normalized by dividing by its mean. On the horizontal axis is shown days since June 10, 2011. A horizontal dashed line indicates the mean of each image's lightcurve and the vertical axis shows the normalized flux values around each curve. For each lightcurve, problematic observations were removed. See Section \ref{sec:obs} for more details. For B1938, the fluxes for images C1 and C2 were added together, and the combined lightcurve is referred to as C. These images were highly blended. In addition, the low signal-to-noise lightcurves of the faintest images in the B0712 and B1938 system are omitted. Lightcurves for all other images are shown individually. Note that B1938 was observed separately from the other two lens systems. Mean fluxes for the lightcurves are listed in Table \ref{VarTab}}
\label{fig:lightcurves_norm_JVLA}
\end{figure}

\begin{table*}
\caption{Variability Test Results}
\label{VarTab}
\begin{tabular}{lccccccccc}
\hline
\footnotesize{Lens}
 & \footnotesize{Image}
 & \footnotesize{Mean}
 & \footnotesize{$\chi_{red}^2$}
 & \footnotesize{Prob.}
 & \footnotesize{$\chi_{red}^2$}
 & \footnotesize{Prob.} 
 & \footnotesize{Mean}
 & \footnotesize{$\chi_{red}^2$}
 & \footnotesize{Prob.} \\
   \footnotesize{ }
 & \footnotesize{ }
 & \footnotesize{Flux}
 & \footnotesize{(Full VLA)$^{\rm a}$}
 & \footnotesize{(Full VLA)$^{\rm b}$}
 & \footnotesize{(VLA A}
 & \footnotesize{(VLA A}
 & \footnotesize{Flux}
 & \footnotesize{(JVLA)$^{\rm d}$}
 & \footnotesize{(JVLA)$^{\rm b}$}\\
  \footnotesize{ }
 & \footnotesize{ }
 & \footnotesize{VLA (mJy)}
 & \footnotesize{ }
 & \footnotesize{ }
 & \footnotesize{Config.$^{\rm c}$)}
 & \footnotesize{Config.$^{\rm b}$)}
 & \footnotesize{JVLA (mJy)}
 & \footnotesize{ }
 & \footnotesize{ }\\

\hline
MG0414 & A1 & \phm{.}189 & 8.73 & $<$0.01 & 2.90 & $<$0.01 & \phm{.}168 & 3.41 & $<$0.01 \\
MG0414 & A2 & \phm{.}167 & 4.16 & $<$0.01 & 1.81 & 0.02 & \phm{.}154 & 3.82 & $<$0.01 \\
MG0414 & B  & 71.4 & 17.5 & $<$0.01 & 1.73 & 0.03 & 62.7 & 0.62 & 85.2\\
MG0414 & C  & 27.7 & 2.73 & $<$0.01 & 0.67 & 83.2 & 25.2 & 1.15 & 30.7\\
B0712 & A+B & 23.8 & 37.1 & $<$0.01 & 7.67 & $<$0.01 & 21.7 & 3.64 & $<$0.01\\
B0712 & C & 5.59 & 4.16 & $<$0.01 & 0.90 & 57.9 & 5.07 & 0.73 & 75.1\\
B0712 & D & 1.20 & 0.88 & 70.4 & 0.57 & 93.1 & 1.13 & 0.14 & $>$99.9\\
B1030 & A & \phm{.}350 & 49.4 & $<$0.01 & 5.98 & $<$0.01 & $^{\rm f}$ &  $^{\rm f}$ & $^{\rm f}$\\
B1030 & B & 29.8 & 20.9 & $<$0.01 & 14.6 & $<$0.01 & $^{\rm f}$ &  $^{\rm f}$ & $^{\rm f}$\\
B1127 & A & 8.50 & 0.85 & 77.5 & 0.45 & 98.2 & $^{\rm f}$ &  $^{\rm f}$ & $^{\rm f}$\\
B1127 & B & 7.22 & 0.99 & 49.9 & 0.30 & 99.9 & $^{\rm f}$ &  $^{\rm f}$ & $^{\rm f}$\\
B1152 & A & 47.9 & 5.67 & $<$0.01 & 2.05 & 0.45 & $^{\rm f}$ &  $^{\rm f}$ & $^{\rm f}$\\
B1152 & B & 15.8 & 3.21 & $<$0.01 & 0.74 & 77.5 & $^{\rm f}$ &  $^{\rm f}$ & $^{\rm f}$\\
B1938 & C & $^{\rm e}$ & $^{\rm e}$ & $^{\rm e}$ & $^{\rm e}$ & $^{\rm e}$ & \phm{.}154 &2.05 & 0.96 \\
B1938 & B & $^{\rm e}$ & $^{\rm e}$ & $^{\rm e}$ & $^{\rm e}$ & $^{\rm e}$ & 30.5 & 0.42 & 97.4 \\
B1938 & A & $^{\rm e}$ & $^{\rm e}$ & $^{\rm e}$ & $^{\rm e}$ & $^{\rm e}$ & 9.20 & 93.3 \\
\hline
\multicolumn{10}{l}{$^{\rm a}$\footnotesize{Reduced $\chi^2$ value calculated when comparing full VLA campaign lightcurve against a line with no variability.}}\\
\multicolumn{10}{p{16cm}}{$^{\rm b}$\footnotesize{Probability of obtaining the given $\chi^2$ value through random chance, assuming a source with no variability, in percentages. While small values imply variation beyond measurement noise, the number of large probabilities measured may indicate overestimation of errors, particularly for the fainter images.}}\\
\multicolumn{10}{l}{$^{\rm c}$\footnotesize{Reduced $\chi^2$ value calculated when comparing only the VLA A configuration lightcurve against a line with no variability. }}\\
\multicolumn{10}{l}{$^{\rm d}$\footnotesize{Reduced $\chi^2$ value calculated when comparing the JVLA campaign lightcurve against a line with no variability.}}\\
\multicolumn{10}{l}{$^{\rm e}$\footnotesize{B1938 was only observed with the JVLA.}}\\
\multicolumn{10}{l}{$^{\rm f}$\footnotesize{B1030, B1127, and B1152 were only observed with the VLA.}}
\end{tabular}
\end{table*}

Errors were calculated as a combination of additive and multiplicative terms. The former is approximated by the RMS error in the residual map after Difmap model fitting has been completed. In addition, a term is needed to quantify the inaccuracy of the modeling. Since the CSOs should have approximately constant fluxes, the scatter in their flux ratios provides an estimate of this uncertainty. We estimate this fractional error to be 0.55\% for the VLA monitoring campaign and 0.45\% for the JVLA campaign.

After the fluxes were calculated, observations with loss or corruption of substantial amounts of data were excised. In addition, observations where all images in a system exhibited simultaneous variations were cut, as well. The latter was judged to occur when the variations could be easily found by eye. In addition, large systematic variations were present towards the end of the B configuration period of the VLA campaign. The relevant times are shown with a shaded region in Figure \ref{fig:lightcurves_norm}. We found no correlations between this region of large variations and the weather, the elevations of the targets, or any other suspected sources. We attempted multiple, independent reductions of the data and performed our analysis using different CSOs as the primary flux calibrator. Still, we were unable to ascertain the cause of these large variations, which appeared for all sources, and were thus unable to remove them. In our subsequent analysis, the data points in this region were not included.

\section{Results}
\label{sec:res}

In this section, we present the results of the radio monitoring campaigns. The normalized, corrected lightcurves for each set of lensed images are shown in Figures \ref{fig:lightcurves_norm} and \ref{fig:lightcurves_norm_JVLA}. As explained in Section \ref{sec:obs}, each lightcurve is divided by the average normalized lightcurves of the CSOs. The resulting lightcurves are then normalized by dividing by their mean fluxes. In addition, images A and B for B0712 and images C1 and C2 of B1938 are combined into composite curves. In both cases, the two images are highly blended. They cannot be resolved from each other, even in the A configuration of the VLA or JVLA. However, this should not be problematic for our analysis as the small image separations imply short time delays ($< 1$ day)\footnote{Moustakas et al., in prep., predict a time delay of 0.04 days between images A and B for B0712.} between the blended images, which are much less than the average time between observations ($\sim 4$ days). 

\begin{figure}
\includegraphics[width=0.47\textwidth]{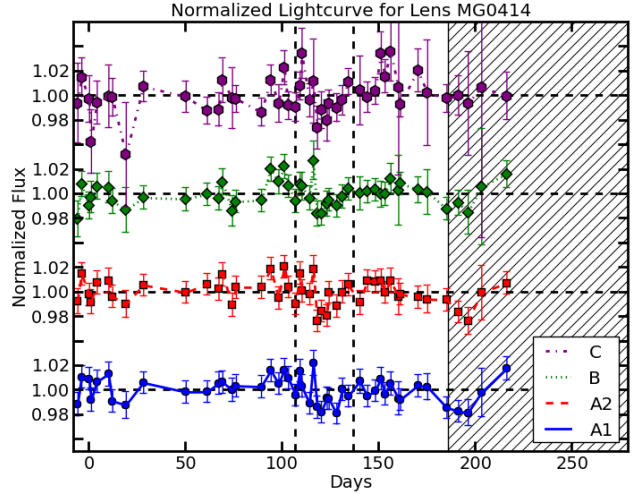}
\caption{
  Lightcurves for MG0414, reduced without baselines greater than 300 kilo-wavelengths, with arbitrary vertical offsets between images. Each lightcurve was normalized by dividing by its mean. On the horizontal axis is shown days since October 21, 2000. A horizontal dashed line indicates the mean of each image's lightcurve and the vertical axis shows the normalized flux values around each curve. In each case, problematic observations were removed. See Section \ref{sec:obs} for more details. Vertical dashed lines indicate the transition from the A to the BnA configuration, and then from the BnA to the B configuration.
}
\label{fig:0414UVR}
\end{figure}

\subsection{Variability}
\label{sec:var}

To quantitatively determine which lensed quasars were variable, we compared each lightcurve to a line of constant flux using a $\chi^2$ test. The results of this test are shown in Table \ref{VarTab}. When testing the entire lightcurves for the VLA observations, only B1127 was consistent with constant flux. However, as mentioned in Section \ref{sec:obs}, most of the VLA lightcurves have large variations near the end of the B configuration phase, which we were unable to remove. The relevant time range is shown with a shaded region in Figure \ref{fig:lightcurves_norm}. Since these variations are unlikely to be intrinsic, these data points were removed from subsequent analysis. A separate $\chi^2$ test was performed for each of the VLA lightcurves using only the A configuration data, the results of which are also shown in Table \ref{VarTab}. These tests find more lightcurves to be consistent with constant flux. These curves are unlikely to have detectable intrinsic variability during this campaign and are thus require further observation to be useful for measuring cosmological parameters. Note, though, that the lack of variation in our campaign does not preclude future seasons of higher variability. As evidenced with observations of B1608, high variability can follow an initial season of small variations in a lensed quasar's flux \citep{fass02}.

Two of the systems, B1030 and B0712, show clear linear trends in flux versus time for most or all images. B1030 shows a steady, monotonic decrease during the campaign, while, for B0712, the combined A+B curve and that for image C both show a monotonic increase with time after a short flat period. These trends are verified by the $\chi^2$ tests, which show that the curves deviate from constant flux at greater than the $3\sigma$ level. This is not true for image D for B0712, but this is likely because of its low signal-to-noise ratio. The trends in these lightcurves are promising signs of intrinsic variability for the two lensed quasars, but the linear time dependences introduce a degeneracy between time delays and magnification that makes it difficult to measure either from these observations alone. 

The MG0414 lightcurve is inconsistent with a constant flux when the entire lightcurve is considered, but it has an unusual shape. It shows an approximately steady flux while the VLA was in the A configuration, an increase in flux while the VLA was in the transitionary BnA configuration, and then an approximately steady flux again while the VLA was in the B configuration. The timing of this increase is suspicious. Such an increase could be produced if the source contains some diffuse emission that is more readily detected in the lower resolution B configuration. In fact, diffuse emission has been detected in the system in the past, with evidence of a jet visible slightly offset from both images A1 and A2 \citep{katz97}. To minimize the contribution from diffuse emission, we reduced the data using only baselines less than 300 kilo-wavelengths, the longest baselines measurable with the B configuration, and uniform weighting. The resulting lightcurves are shown in Figure \ref{fig:0414UVR}. Since the rise in flux at the array configuration change is no longer visible, it was likely a result of diffuse emission and not intrinsic variation of the source. 

In order to perform the {\it difmap} reduction of the MG0414 VLA data, we had to use a variable position model, unlike the other systems\footnote{Variations in fit image positions for the MG0414 system were $\sim 0.1 ''$. Three outliers with substantially larger variations from the mean positions were thrown out.}. Fixed position models failed to separate the fluxes of images A1 and A2 in the B configuration. This is likely a result of the contribution to the flux from the jets. Variations in the flux ratio between the main images and the jets could create different centroids for the image/jet combination. While this creates additional uncertainty in the resultant fluxes, more complex models, taking the jets into account, failed. These models introduced large, likely erroneous variations in flux, implying that more free parameters were introduced than could be constrained with the available information. Large uncertainties in the location of the source of variability would create consequently large uncertainties on the time delay and cosmological parameters \citep{barn14}.

While the B1938 lightcurves do not show any obvious trends, the $\chi^2$ test shows that the lightcurve for the composite image C is inconsistent with constant flux at a high significance. While the other, fainter images were not inconsistent with constant flux, this result does suggest that this system may be intrinsically variable, and it may be possible to measure a time delay in the future. In addition, the JVLA lightcurves of the brightest MG0414 and B0712 images were also inconsistent with constant flux at high significance, while the fainter lightcurves were not. In all three cases, the non-detection of variability in the fainter images could be due to lower signal to noise ratios. 

These cases are similar to the observations of B1608. As shown in \citet{fass02}, B1608 varied relatively little in the first season of their three-season monitoring campaign, with subsequently more variations in time. However, the first season of this B1608 data is significantly variable according to our variability test, at a $> 99.99$ \%level. Similarly, while the B0712 JVLA data lacks obvious features, the VLA data of the same system was highly variable. So, while our B1938 data showed no obvious trends with time, further observations could yield more interesting variations. The same could be true for any of the other systems that showed little variation during our campaigns. 

\begin{figure*}
\includegraphics[width=\textwidth]{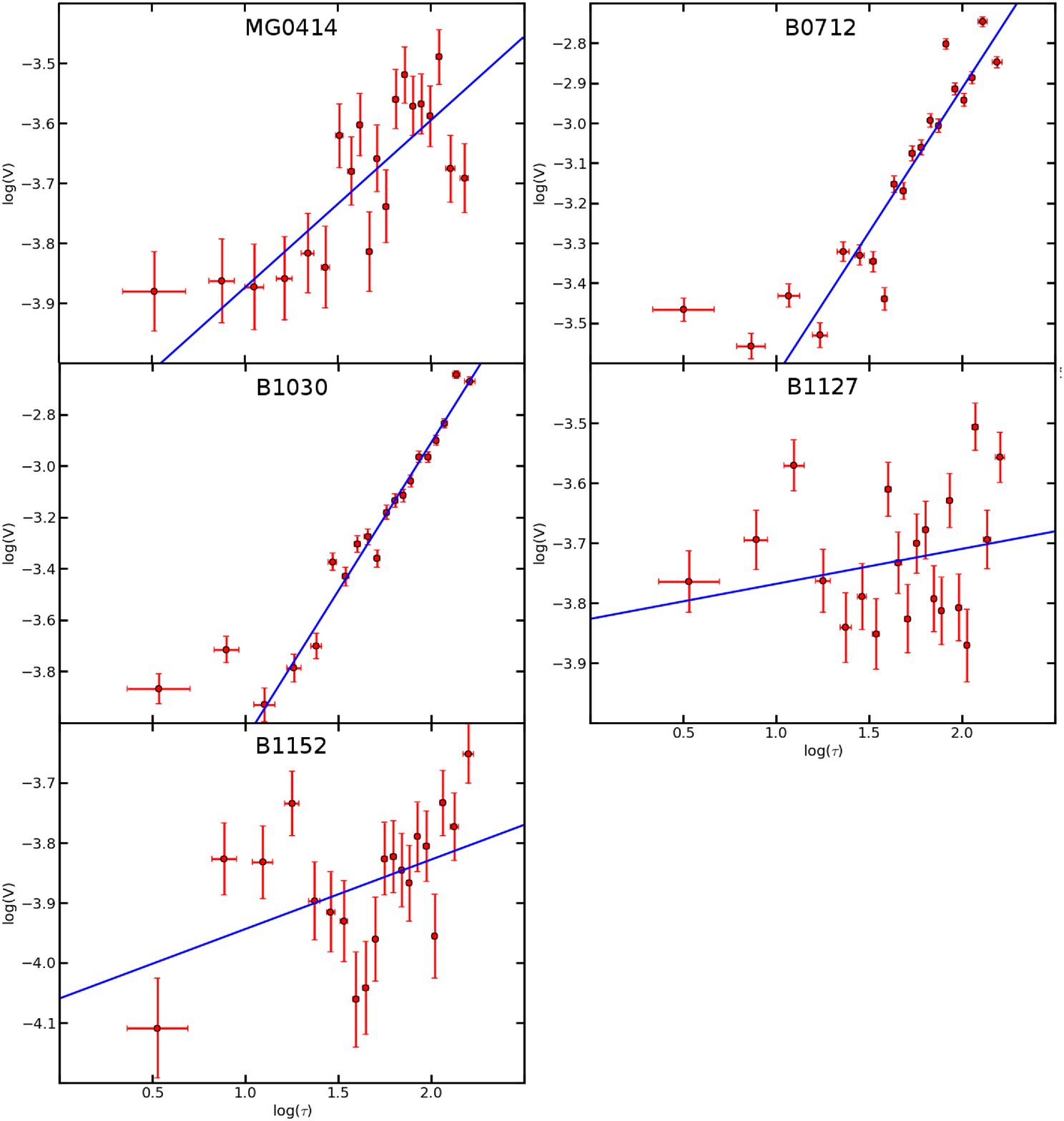}
\caption{
  Results of structure function analysis are shown for the brightest image of each lensed radio source observed with the VLA (See Sections \ref{sec:SF} for details). Fits used only points for which $\tau > 10$ days.
}
\label{SFPlot}
\end{figure*}

\begin{figure*}
\includegraphics[width=\textwidth]{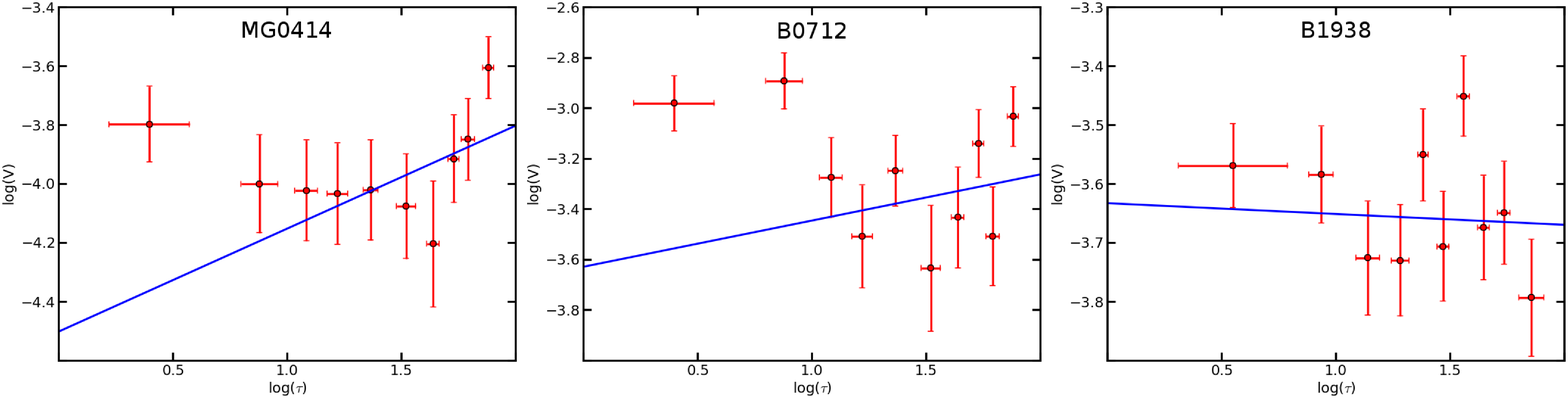}
\caption{
  Results of structure function analysis are shown for the brightest image of each lensed radio source observed with the JVLA (See Sections \ref{sec:SF} for details). Fits used only points for which $\tau > 10$ days.
}
\label{SFPlotJVLA}
\end{figure*}

\begin{centering}
\begin{table}
\caption{Structure Function Analysis Results}
\label{SFTab}
\begin{tabular}{lcc}
\hline
\footnotesize{Lens}
 & \footnotesize{Image}
 & \footnotesize{SF Power}\\
\footnotesize{}
 & \footnotesize{}
 & \footnotesize{Law}\\
\hline
\multicolumn{3}{l}{VLA Campaign} \\
\hline
MG0414 & A1\phm{+B)} & \phm{-}0.28 $\pm$ 0.07\\
MG0414 & A2\phm{+B)} & \phm{-}0.12 $\pm$ 0.06\\
MG0414 & B\phm{(+B)} & \phm{-}0.84 $\pm$ 0.06\\
MG0414 & C\phm{(+B)} & \phm{-}0.25 $\pm$ 0.06\\
B0712 & (A+B) & \phm{-}0.72 $\pm$ 0.07\\
B0712 & C\phm{(+B)} & \phm{-}0.95 $\pm$ 0.09\\
B0712 & D\phm{(+B)} & \phm{-}0.05 $\pm$ 0.12\\
B1030 & A\phm{(+B)} & \phm{-}1.15 $\pm$ 0.05\\
B1030 & B\phm{(+B)} & \phm{-}0.58 $\pm$ 0.08\\
B1127 & A\phm{(+B)} & \phm{-}0.06 $\pm$ 0.09\\
B1127 & B\phm{(+B)} & \phm{-}0.16 $\pm$ 0.06\\
B1152 & A\phm{(+B)} & \phm{-}0.12 $\pm$ 0.08\\
B1152 & B\phm{(+B)} & \phm{-}0.73 $\pm$ 0.09\\
\hline
\multicolumn{3}{l}{JVLA Campaign} \\
\hline
MG0414 & A1\phm{+B)} & \phm{-}0.35 $\pm$ 0.22\\
MG0414 & A2\phm{+B)} & \phm{-}0.83 $\pm$ 0.25\\
MG0414 & B\phm{(+B)} & -0.30 $\pm$ 0.32\\
MG0414 & C\phm{(+B)} & -0.64 $\pm$ 0.27\\
B0712 & (A+B) & \phm{-}0.18 $\pm$ 0.29\\
B0712 & C\phm{(+B)} & \phm{-}0.77 $\pm$ 0.11\\
B0712 & D\phm{(+B)} & -0.41 $\pm$ 0.26\\
B1938 & C\phm{(+B)} & -0.02 $\pm$ 0.19\\
B1938 & B\phm{(+B)} & -0.00 $\pm$ 0.17\\
B1938 & A\phm{(+B)} & -0.14 $\pm$ 0.21\\
\hline
\end{tabular}
\end{table}
\end{centering}

\subsubsection{Structure Function Analysis}
\label{sec:SF}

To further characterize the variability of the lightcurves, we used the first-order structure function (SF). The SF provides information on the variability and fluctuation modes of lightcurves. The SF is defined by \begin{equation}
V(\tau) = \frac{1}{N}\sum^{N}_{i=1}\left[S\left(t_i\right)-S\left(t_i+\tau\right)\right]^2
\end{equation}
where $S\left(t_i\right)$ are the $N$ measured fluxes in the overlap region. The SF provides information on the temporal variability of lightcurves, including the time scales of variation and the fluctuation modes (i.e. flickering, shot-noise) \citep{simon85}. 

We measured the SFs for our lightcurves by binning pairs of flux points based on temporal separation $\tau$. The SFs for the brightest image of each lensed radio source are plotted in Figures \ref{SFPlot} and \ref{SFPlotJVLA}, with $\tau$ binned into 20 equal size bins for the sources observed with the VLA and 10 bins for the sources observed with the JVLA (due to the shorter campaign length). 

For a radio source that is varying with time, we would expect the SF to be constant at small time scales where noise dominates the signal, followed by a power law increase and then a plateau or drop off on time scales over which the lightcurve is no longer correlated \citep{gupta08,abdo10}. To test which images were variable, we fit a function of the form $V \propto \tau^{\alpha}$ to each image's SF, using only points for which $\tau > 10$ days. Time differences below 10 days begin to approach the observation cadence and are dominated by noise, which is apparent for the plots of B0712 and B1030 in Figure \ref{SFPlot}. Noise becomes exponentially less important as $\tau$ increases, so excising this lower plateau region should have a negligible effect on the power law fit. Results did not differ significantly when the value of the cutoff was varied. The results of these fits are shown in Table \ref{SFTab}. SF power laws significantly different from zero suggest a variable lightcurve. 

The results of this test are similar to those found using the $\chi^2$ test. B0712 and B1030, which showed linear trends of flux with time, had significantly non-zero SF power laws, except for B0712D, which had a low signal to noise ratio. Similarly, three out of the four images of MG0414 had power laws that differed from zero by at least a 3$\sigma$ level. The other lightcurves from the VLA campaign that failed the $\chi^2$ test fail this one as well, with the exception of B1152B. However, since the brighter image A failed the test, it is unlikely that real variability would be detected with a lower signal-to-noise ratio. 

Additionally, the lack of a drop off for most of the sources could indicate that the characteristic timescales are longer than the campaign lengths \citep{gupta08}. The SFs measured then may not be characteristic of the sources in the long term. This is especially true for the JVLA lightcurves. None of the SF power laws were significantly different from zero for these sources, despite the fact that the brightest image from each passed the $\chi^2$ test. This could be because the campaigns were not long enough to precisely measure the SFs. However, the general agreement between the two tests for the VLA sources does offer the previous results some validity.

\begin{figure*}
\includegraphics[width=\textwidth]{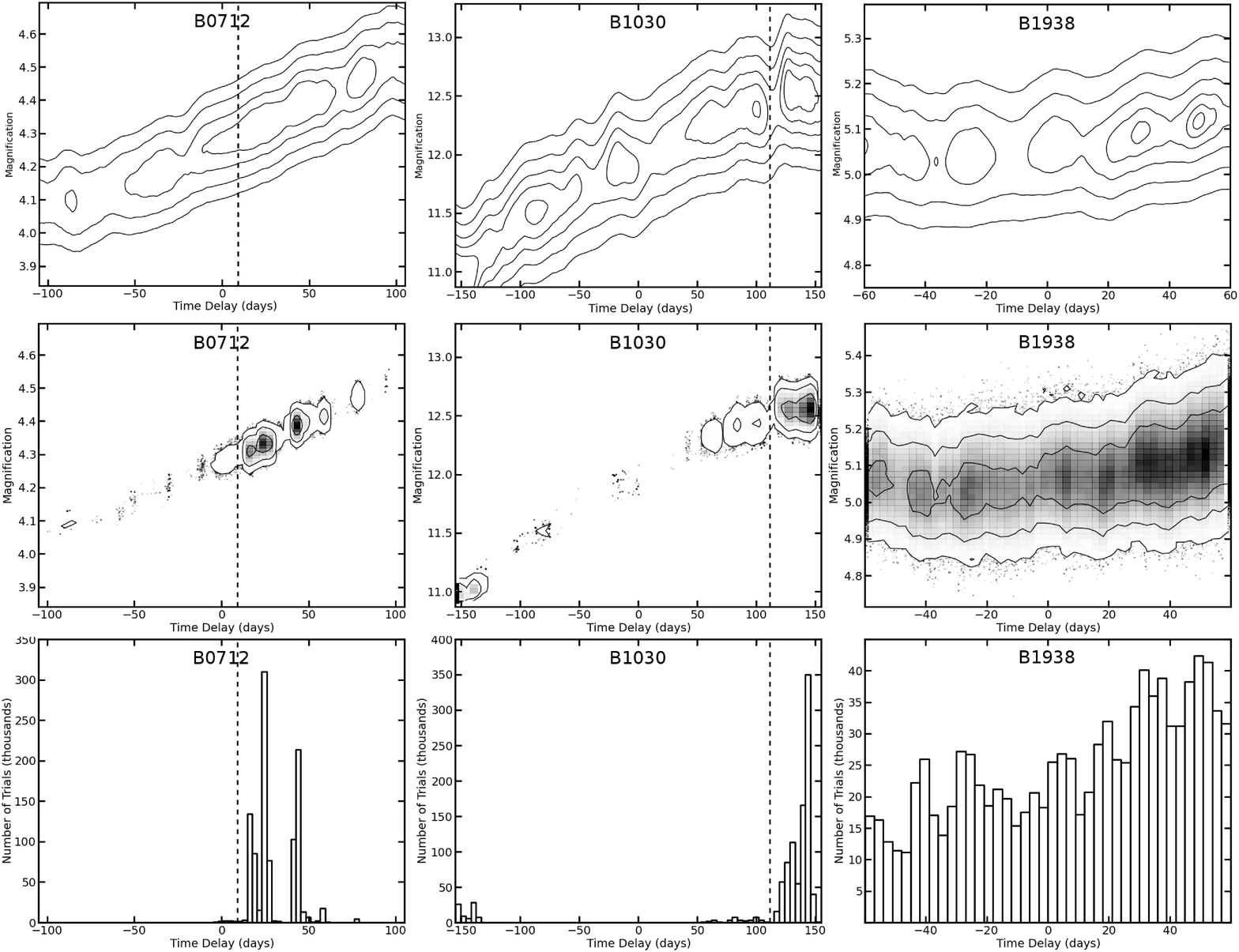}
\caption{
  Contour plots displaying the results of time delay measurements are shown for B0712, B1030, and B1938 between the images pairs C and A+B, B and A, and B and C, respectively. A positive time delay here means that the image that is expected to vary first does so (images C, A, and B for B0712, B1030, and B1938, respectively). In the top plots, contours are constructed from the $D_{4,2}$ dispersion metric using $\delta=10.5$\ days. The contour levels are arbitrary, meant only to illustrate the general shapes and minima, as well as the degeneracies in the time delay and magnification parameters. In the middle plots, the results of MCMC trials based on a $\chi^2$ method are shown for B0712, B1030, and B1938. The contours contain 68.3\%, 95.4\%, and 99.7\% of the trial points. The bottom plots are the histograms of the marginalized probability distribution functions for the time delays measured using the MCMC method. The vertical dashed lines indicate the predicted time delays of 9 and 111 days between the relevant B0712 and B1030 images, respectively (Moustakas et al., in prep., \citealt{xan98}). The delay of 153$^{+29}_{-57}$ predicted by \citet{saha06} is beyond the range of time delays we could reasonably test with our campaign length.
}
\label{ConPlot}
\end{figure*}

\begin{figure*}
\includegraphics[width=\textwidth]{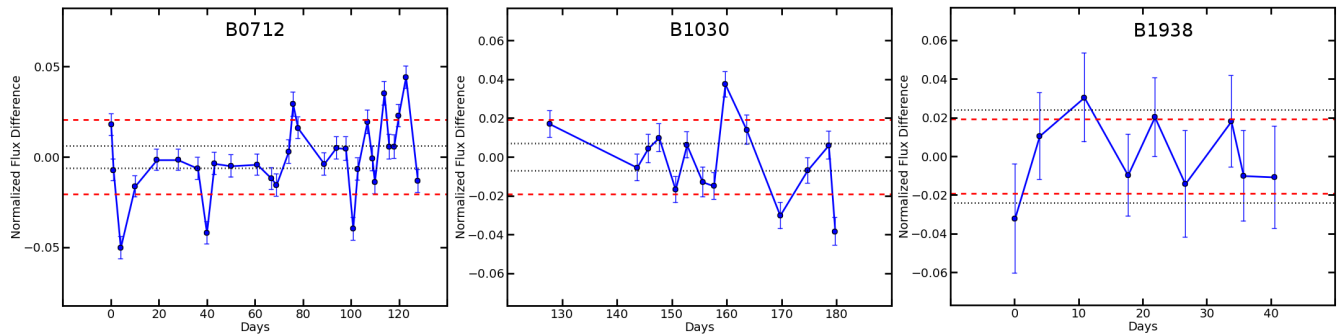}
\caption{
  Plots were constructed based on the time delay-magnification pair with the lowest dispersion for each image pair we analyzed in Section \ref{sec:TDA}. In each case, the brighter image lightcurve was multiplied by the magnification, offset by the time delay, and interpolated onto the fainter image lightcurve. The difference between the lightcurves was taken and this difference lightcurve, divided by the mean flux of the fainter image lightcurve, is plotted here, with errors calculated based on the uncertainties of two input lightcurves. The dotted lines represent the expected dispersion based purely on measurement errors, while the dashed line is the RMS variability of the difference lightcurve.
}
\label{sub_lc_plot}
\end{figure*}

\begin{figure}
\includegraphics[width=0.48\textwidth]{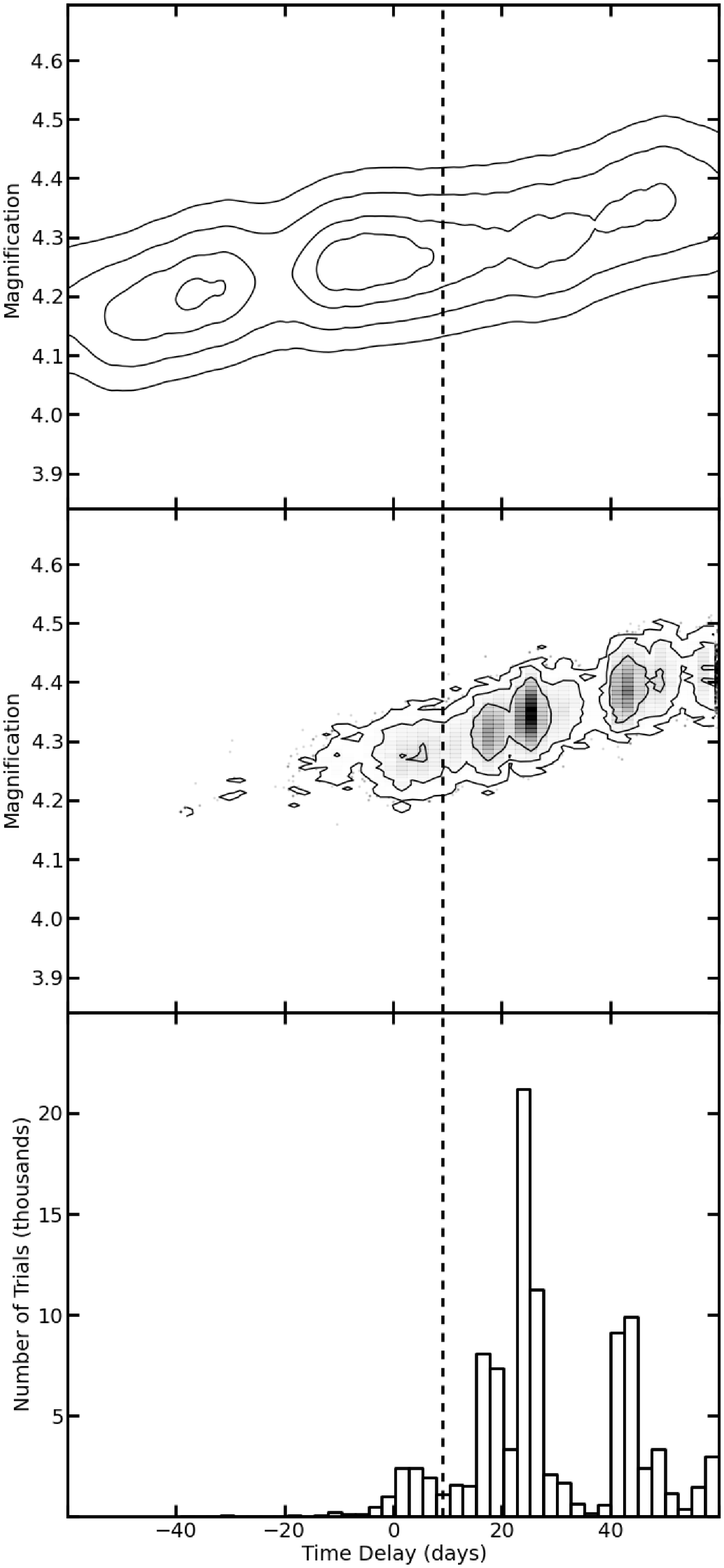}
\caption{
  Contour plots for B0712, jointly fitting the C and A+B lightcurves from both campaigns. A positive time delay here means that the fainter image C is varying first. In the top plot, contours are constructed from the $D_{4,2}$ dispersion metric using $\delta=10.5$\ days. Note that the choice of dispersion metric and $\delta$ value had little effect on our analysis. The contour levels are arbitrary, meant only to illustrate the degeneracies in the time delay and magnification parameters. The middle plot shows the results of MCMC trials based on a $\chi^2$ method. The contours contain 68.3\%, 95.4\%, and 99.7\% of the trial points. The bottom plot is a histogram of the marginalized probability distribution function for the time delays measured using the MCMC method. The vertical dashed lines indicate the predicted time delay of 9 days between the relevant B0712 images (Moustakas et al., in prep.)
}
\label{ConPlot_0712JF}
\end{figure}

\subsection{Flux Ratios}

While some of the strong lenses we observed could not be used to measure time delays due to lack of variability, their lightcurves are still useful for measuring flux ratios between the images. Flux ratios were measured by comparing the mean fluxes of different lightcurves. The results, calculated without time shifts, are shown in Table \ref{FRTab}. For B1127 and B1152, we measured the flux ratios using both the full lightcurves and using only the less variable A configuration data. For both lenses, there was little difference between these ratios. For MG0414, the lightcurves started to vary significantly approximately when the VLA switched from the A to the BnA configuration, so we did not attempt to measure the flux ratios for the B or BnA configuration data. 

The statistical uncertainties on these flux ratio measurements were very small $\left(\sim 10^{-5}\right)$. We would expect the dominant source of error to come from misalignment of the lightcurves due to the unknown time delays. To estimate this source of systematic error, we introduced a range of temporal offsets between the images' lightcurves, measuring the flux ratio in the overlap region for each case. We took the range of the flux ratios measured in this way as an estimate of the systematic uncertainty, which is shown in Table \ref{FRTab}.  

Because the MG0414 VLA lightcurves and all of the JVLA lightcurves showed some signs of variation, the flux ratios measured for their images are less reliable than the others. Still, the MG0414, B0712, and B1938 flux ratios are all consistent with earlier measurements \citep{katz93,king97,jack98}. In addition, our flux ratio for B1152 is consistent with that measured by \citet{myers99}. However, while the lightcurve for the combined image C of B1938 was significantly variable according to the test described in Section \ref{sec:var}, we found the magnification to only weakly depend on the time delay in our analysis (See Figure \ref{ConPlot}). Even with the uncertainties in the measurement of these flux ratios, they could prove useful for future monitoring campaigns of these lenses. 

\subsection{Time Delay Analysis}
\label{sec:TDA}

For B0712, B1030, and B1938, the three systems that displayed signs of intrinsic variability excluding the unreliable MG0414 data (see Section \ref{sec:var}), we attempted to measure time delays between the brightest images. To do so, we first used the dispersion method used by \citet{fass02} based on the method of \citet{pelt94,pelt96}. We used the D$_2$ and D$_{4,2}$ metrics, the latter of which is characterized by a width parameter, $\delta$. We minimized the dispersion by calculating the values of these metrics at each point of a grid of time delay and magnification values for each image pair for D$_2$ and D$_{4,2}$, using a range of values for $\delta$. Contour plots of the D$_{4,2}$ metric using $\delta = 10.5$\ days are shown in Figure \ref{ConPlot} in the top portion\footnote{We found similar results using the D$_2$ and D$_{4,2}$ metrics and with different values of $\delta$.}. The contour levels are arbitrary and are meant only to show the shape of the distribution. The degeneracies between the time delay and magnification are illustrated in this plot for B0712 and B1030. However, it is difficult to measure the uncertainties of time delays based on this dispersion method alone. 

To estimate the uncertainties on the time delay measurements, we also measured time delays using a Monte Carlo Markov Chain (MCMC) method based on a $\chi^2$ metric. To calculate $\chi^2$, we interpolated one of the lightcurves being compared onto the other, after adding a time delay and magnification. The order of interpolation was then reversed, and the $\chi^2$ was taken as the median of the two values. The results of this method are shown in the middle portion of Figure \ref{ConPlot}, and marginalized probability distribution functions for the time delays are shown in the bottom portion. In this case, the contours represent the regions containing 68.3\%, 95.4\%, and 99.7\% of the trial points. Note that the shapes of the contours and positions of extrema are similar for both time delay measurement methods, indicating the robustness of our measurements. Our confidence regions are large compared to the campaign lengths ($>$ 50 days), with a non-negligible distribution across most of the positive range for the time delays in each case. As can be expected from the degeneracies and quality of data, the uncertainties are too large to state a robust measurement of any of these time delays, although the results do seem to confirm the expected sign of the delays. 

As an additional check of the time delays and possibly for the presence of extrinsic variability, we compared the image lightcurves using the time delay-magnification pair with the lowest dispersion value. For each lens system, we multiplied the brighter image lightcurve by the relevant magnification, offset it by the corresponding time delay, and interpolated it onto the fainter image lightcurve. We then subtracted these two lightcurves, and these difference lightcurves, divided by the mean flux of the fainter image, are plotted in Figure \ref{sub_lc_plot}. In each plot, the dotted line represents the expected dispersion based purely on the measurement errors, while the dashed line corresponds to the RMS variability of the difference lightcurve. For B1938, these two lines are nearly coincident, while they differ significantly for B0712 and B1030\footnote{As in Section \ref{sec:var}, comparison to zero yields a reduced $\chi^2$ value of 0.6 for the B1938 difference lightcurve and reduced $\chi^2$ values greater than 7 for both B0712 and B1030.}. For the correct time delay-magnification pair, in the absence of extrinsic variations, we would expect agreement with a line of zero flux, as deviation of the difference lightcurve from zero should be purely governed by scatter characterized by the measurement errors. While the agreement between the measurement errors and RMS variability does not imply that the B1938 time delay is correct, the lack of agreement for the other systems does have implications. There may be several causes for the discrepancy. The time delay-magnification pairs could be incorrect. The measurement errors could also be underestimated or there could have been a calibration error (which could also lead to underestimated errors). Another possibility is the presence of extrinsic variability in the lightcurves. Although radio observations are less prone to microlensing than in the optical range, they are not immune. In addition, galactic scintillation could cause such variation. \citet{koop03} found extrinsic variations of up to $\sim 40$\% peak-to-peak for flux ratios of a sample of lenses that included B0712. While there are a number of possibilities for the discrepancies in the difference lightcurves for B0712 and B1030, the degeneracy between the time delay and magnifications makes it difficult to differentiate between the causes.

Despite the time delay-magnification degeneracy, there is further information available for B0712 in the additional JVLA campaign. While the MG0414/B0712 JVLA campaign block was admittedly short ( it contains only 14 data points), we performed joint time delay fits using both the VLA and JVLA lightcurves for B0712 and using both methods described above. The results are shown in Figure \ref{ConPlot_0712JF} in the same manner as Figure \ref{ConPlot}. Both contour plots have similar shapes, and the degeneracy between the parameters is still visible, although the magnification information from the JVLA campaign has reduced it somewhat. However, the confidence region is still large compared to the size of the campaigns, so we still cannot robustly determine the time delay. Additional observations are necessary to measure a time delay for B0712, as well as for B1030 or B1938. 

\begin{table*}
\caption{Flux Ratios}
\label{FRTab}
\begin{tabular}{lccccc}
\hline
\footnotesize{Lens}
 & \footnotesize{Image}
 & \footnotesize{Flux Ratio}
 & \footnotesize{Uncertainty}
 & \footnotesize{Flux Ratio}
 & \footnotesize{Uncertainty}\\
\footnotesize{}
 & \footnotesize{Pair}
 & \footnotesize{(entire campaign)}
 & \footnotesize{(entire campaign)}
 & \footnotesize{(A Config. only)}
 & \footnotesize{(A Config. only)}\\
\hline
\multicolumn{6}{l}{VLA Campaign} \\
\hline
MG0414 & A1/B & $^{\rm a}$ & $^{\rm a}$ & 2.68 & 0.14\\
MG0414 & A2/B & $^{\rm a}$ & $^{\rm a}$ & 2.39 & 0.12\\
MG0414 & \phm{1}C/B & $^{\rm a}$ & $^{\rm a}$ & 0.39 & 0.02\\
B1127 & \phm{1}A/B & 1.18 & 0.01 & 1.19 & 0.03\\
B1152 & \phm{1}A/B & 3.03 & 0.13 & 3.05 & 0.17\\
\hline
\multicolumn{6}{l}{JVLA Campaign} \\
\hline
MG0414 & A1/B & $^{\rm b}$ & $^{\rm b}$ & 2.69 & 0.04\\
MG0414 & A2/B & $^{\rm b}$ & $^{\rm b}$ & 2.45 & 0.04\\
MG0414 & \phm{1}C/B & $^{\rm b}$ & $^{\rm b}$ & 0.40 & 0.01\\
B0712 & (A+B)/C & $^{\rm b}$ & $^{\rm b}$ & 4.26 & 0.21\\
B0712 & \phm{1}D/C & $^{\rm b}$ & $^{\rm b}$ & 0.21 & 0.02\\
B1938 & \phm{1}A/B & $^{\rm b}$ & $^{\rm b}$ & 0.30 & 0.01\\
B1938 & \phm{1}C/B & $^{\rm b}$ & $^{\rm b}$ & 5.05 & 0.10\\
\hline
\multicolumn{6}{l}{$^{\rm a}$\footnotesize{The MG0414 lightcurves can only be reasonably approximated as constant when the VLA was in the A configuration.}}\\
\multicolumn{6}{l}{$^{\rm b}$\footnotesize{The entirety of the JVLA campaign was in the A configuration.}}
\end{tabular}
\end{table*}

\section{Conclusions and Future Work}
\label{sec:dis}

We carried out radio monitoring campaigns of six strongly lensed radio sources using the VLA and JVLA: MG 0414+0534, CLASS B0712+472, JVAS B1030+074, CLASS B1127+385, CLASS B1152+199, and B1938+666. A primary purpose of the campaigns was to locate suitably variable sources, with variations that can be reasonably measured, to use for estimation of cosmological parameters. The lightcurves from our campaigns exhibited a range of degrees of variation.

\begin{itemize}


\item{The VLA lightcurves of MG0414, B0712, and B1030 passed both our $\chi^2$ and structure function variability tests at the 99.7\% level, while only the brightest images of the JVLA lightcurves passed only the $\chi^2$ test at the 95.4\% level. However, the JVLA lightcurves were probably too short for the SF analysis.}


\item{Variation in the MG0414 VLA lightcurves was likely spurious, caused by diffuse emission preferentially detected in the VLA B configuration. The variation largely disappeared when long baselines were cut, and there are known jets next to the brightest images \citep{katz97}.}

\item{We attempted to measure time delays for B0712, B1030, and B1938. We used both a dispersion-based grid search method and a $\chi^2$-based MCMC method, finding similar results for both. The uncertainties on the parameters from the MCMC method were large for all three systems, with the uncertainty on the time delay large compared to the length of the campaigns. Therefore, we were unable to measure time delays with any precision for our sample, even with information from both campaigns for B0712.}

\end{itemize}

Despite our inability to measure time delays, the results of our campaign will be useful for future work. Although B0712 and B1030 showed linear flux trends during the VLA campaign, which make measurement of a time delay difficult using our data alone, the strong time variations make these promising candidates for follow-up observations. Even if these systems are observed during lulls in variability, the data will be useful to break the time delay-magnification degeneracy that currently plagues them. While the JVLA B0712 observations were unable to meaningfully break this degeneracy, the campaign was exceptionally short. We were only able to obtain fourteen good observations of B0712 with the JVLA over a period of $\sim$ 90 days. 

In addition, the B1938 lightcurve passed our variability test, although it showed no clear trends with time and we were unable to measure a time delay for it. This is similar to the first season of data \citet{fass02} obtained for B1608. Radio-loud active galactic nuclei can go through periods of little activity followed by large variations, as evidenced by the subsequent seasons of observation \citet{fass02} obtained for B1608. With additional monitoring campaigns, B1938 could exhibit a similar increase in variability, allowing a precise measurement of the system's time delays. The same could hold true for B1127, B1152, or MG0414. Observations at a different frequency may be able to overcome the latter's problem with diffuse emission. In conclusion, most of the systems observed should be targeted with followup monitoring campaigns, especially B0712 and B1030.

\bigskip

CDF acknowledges support from the National Science Foundation collaborative grant ``Collaborative Research: Accurate cosmology with strong gravitational lens time delays'' (AST-1312329).

LVEK is supported in part through an NWO-VICI career grant (project number 639.043.308).


\bibliographystyle{mn2e}

\bibliography{rum}

\end{document}